\documentclass[preprint,12pt]{elsarticle}

\usepackage{amssymb}
\usepackage{amsmath}
\usepackage{pifont}
\usepackage[labelfont=bf]{caption}
\usepackage{xcolor}
\usepackage[colorlinks=true]{hyperref}
\usepackage{subcaption}
\usepackage{multirow}
\usepackage{float}

\journal{Image and Vision Computing}

\begin{document}

\begin{frontmatter}



\title{FreqNet: A Frequency-domain Image Super-Resolution Network with Discrete Cosine Transform}


\author[inst1]{Runyuan Cai}

\affiliation[inst1]{organization={Department of Computer Science and
Engineering},%Department and Organization
            addressline={Shanghai Jiao Tong University}, 
            city={Shanghai},
            postcode={200240}, 
            country={China}}

\author[inst2]{Yue Ding}
\author[inst1]{Hongtao Lu \corref{cor1}}
\cortext[cor1]{Corresponding author}

\affiliation[inst2]{organization={School of Software},%Department and Organization
            addressline={Shanghai Jiao Tong University}, 
            city={Shanghai},
            postcode={200240}, 
            country={China}}

\begin{abstract}
Single image super-resolution(SISR) is an ill-posed problem that aims to obtain high-resolution (HR) output from low-resolution (LR) input, during which extra high-frequency information is supposed to be added to improve the perceptual quality. Existing SISR works mainly operate in the spatial domain by minimizing the mean squared reconstruction error. Despite the high peak signal-to-noise ratios(PSNR) results, it is difficult to determine whether the model correctly adds desired high-frequency details. Some residual-based structures are proposed to guide the model to focus on high-frequency features implicitly. However, how to verify the fidelity of those artificial details remains a problem since the interpretation from spatial-domain metrics is limited. In this paper, we propose FreqNet, an intuitive pipeline from the frequency domain perspective, to solve this problem. Inspired by existing frequency-domain works, we convert images into discrete cosine transform (DCT) blocks, then reform them to obtain the DCT feature maps, which serve as the input and target of our model. A specialized pipeline is designed, and we further propose a frequency loss function to fit the nature of our frequency-domain task. Our SISR method in the frequency domain can learn the high-frequency information explicitly, provide fidelity and good perceptual quality for the SR images. We further observe that our model can be merged with other spatial super-resolution models to enhance the quality of their original SR output.

\end{abstract}


% \begin{highlights}
% \end{highlights}

\begin{keyword}
Single Image Super-Resolution \sep Frequency Domain \sep Deep Learning
\PACS 0000 \sep 1111
\MSC 0000 \sep 1111
\end{keyword}

\end{frontmatter}


\section{Introduction}
\label{sec:intro}
Single image super-resolution(SISR) aims to recover high-frequency details for a high-resolution(HR) image from one of its degraded low-resolution(LR) version. After years of development, the SISR has been widely used in many computer vision tasks, such as media content enhancement\cite{2-1}, medical imaging\cite{1} and satellite imaging\cite{2}. Traditional state-of-the-art SR methods mainly adopt the example-based\cite{2-1} strategy, exploiting internal similarities or learning a mapping from the external dictionary. The sparse-coding-based SR\cite{2-2} is one of the most representative methods.

Recently, deep convolutional neural network (CNN) based SISR methods have achieved significant improvements over traditional methods. Deep learning-based methods treat this problem as a dense image regression task, which learns an end-to-end image mapping function represented by a CNN between LR and HR images. Dong et al.\cite{3} proposed SRCNN that first adopted deep learning into SISR using a three-layer CNN to represent the mapping function. Residual block\cite{7} was later introduced into SISR in SRResNet\cite{6} and improved in EDSR\cite{8}. Residual block makes it possible to build deeper or wider networks. Zhang et al.\cite{9} and Tong et al.\cite{10} adopted dense blocks\cite{11} to combine features from different levels. Zhang et al.\cite{12} improved residual block by adding channel attention. Based on the progress of non-blind methods, blind super-resolution methods\cite{2-3}, which aim at complex degradation models in real scenarios, have received increasing attention recently.

The SISR methods mentioned above commonly use the minimization of the mean squared error (MSE) between the recovered SR image and the HR ground truth as the optimization target. Minimizing spatial MSE also maximizes the peak signal-to-noise ratio (PSNR), which is a common measure used to evaluate SR algorithms. However, such a pipeline often results in blurry effects because the high-frequency textures have been excessively destructed in the degrading process and are hard to predict. Generative adversarial networks (GANs)\cite{19} based SISR approaches are proposed to relieve the above problems. However, the unpleasant hallucinations and artifacts caused by GANs further pose more challenges. Zhang et al.\cite{12} further proposed a residual-in-residual (RIR) structure to bypass the redundant low-frequency information through multiple skip connections, implicitly guiding the network to focus on learning high-frequency information. However, since the commonly used PSNR and structural similarity index measure(SSIM) are based on per-pixel loss and picture global information, respectively, their perception of high-frequency details is limited. To the best of our knowledge, current spatial domain-based methods do not have an explicit approach for learning high-frequency information and verifying the fidelity of output artificial details.

To practically resolve this problem, we propose FreqNet, a frequency-domain-based super-resolution network, to directly learn the reconstruction of high-frequency features. The proposed network contains two parallel flows: the Spatial Extraction Network(SEN) and the Frequency Reconstruction Network(FRN), in order to make use of both domains' information. We first convert both LR and HR images to frequency coefficients using discrete cosine transform (DCT)\cite{13}, then reshape them to obtain DCT feature maps. The SEN takes standard LR images as input, through the spatial feature reconstruction trunk and the down-sampling shrinking trunk to obtain one component of target HR DCT feature maps. The FRN is purely operated on frequency domain, which takes LR DCT feature maps as input, through the frequency-domain reconstruction trunk to obtain the other component. The weighted sum of two components makes our final frequency domain output, which can be converted to SR image through inverse discrete cosine transform(iDCT)\cite{13}. Thanks to the characteristic of DCT, we can easily merge our output with any other SR model to enhance the high-frequency details of its output. We further propose depth-wise residual block(DWRB) and deformable residual block(DRB) to be implemented respectively in FRN and SEN that can better use the characteristics of the frequency domain feature maps. As the ability of spatial MSE (and PSNR) to capture high-frequency detail is very limited, we propose a frequency-domain loss function to evaluate the quality of the output SR image.

Overall, our contributions are three-fold: (1) We propose FreqNet, a frequency-domain-based SISR network, to learn the high-frequency features explicitly with a specially designed pipeline. Our network can produce perceptually satisfying results with high-fidelity details. (2) We propose depth-wise residual block structure and deformable residual block structure to fit the nature of frequency-domain feature extraction. Both structures can improve our network's reconstruction quality and feature extraction ability. (3) We propose a frequency-domain loss function and a corresponding metric that measures output quality from the accuracy of high-frequency detail reconstruction.

\section{Related Works}

Numerous image deep learning-based SR methods have been studied in the computer vision community. Here we focus on works related to CNN-based methods and the works on frequency-domain learning.

\subsection{Image Super-Resolution with CNN}

Numerous methods have proven the effectiveness of the CNN-based pipeline on image super-resolution tasks. The pioneering work was done by Dong et al.\cite{3}, their proposed SRCNN for image SR achieved superior performance against previous works. Kim et al. proposed VDSR\cite{4} and DRCN\cite{5} by introducing residual learning to ease the training difficulty and significantly improve accuracy. Tai et al. introduced recursive blocks in DRRN\cite{15} and memory blocks in MemNet\cite{16}. A faster network structure FSRCNN\cite{17} was proposed to accelerate the pipeline of SRCNN. Ledig et al.\cite{6} introduced ResNet\cite{7} to construct a deeper network, SRResNet, for image SR. They also proposed SRGAN with perceptual losses\cite{18} and generative adversarial network (GAN)\cite{19} for photo-realistic SR. Such GAN based model was then introduced in ESRGAN\cite{20}, which confirmed that dropping the batch normalization layers can result in better performance. Although SRGAN and ESRGAN can alleviate the blurring and over smoothing artifacts, their predicted results may not be faithfully reconstructed and produce unpleasing artifacts. By removing unnecessary modules in conventional residual networks, Lim et al.\cite{8} proposed EDSR and MDSR, which achieve significant improvement. Zhang et al.\cite{12} introduce channel attention to residual block. Blind-SR methods have also received increasing attention recently, aiming at complex degradation models in real scenarios by estimating degradation kernel using an extra module\cite{2-5}. The methods of \cite{2-4},\cite{2-6} and \cite{2-7} achieved state-of-the-art performance in real-world scenario with multiple modelling strategies.

However, all these CNN-based methods operate on the spatial domain. The information on the frequency domain is not directly used, though the recovery of high-frequency information is precisely the target of the image super-resolution task. 

\subsection{Frequency-Domain based Deep Learning}

Projecting image to frequency domain provides a new perspective for various computer vision tasks. Remarkable performance has been achieved in some frequency-domain works. Works of \cite{21} ,\cite{22} and \cite{23} jointly train auto-encoder-based networks on compression and inference tasks with frequency-domain input. \cite{24} extracts features from the frequency domain to classify images. \cite{25} proposes a model conversion algorithm to convert the spatial-domain CNN models to the frequency domain. \cite{14} propose a method of learning in the frequency domain using DCT-based sparse image representations, proving that we can use frequency-domain information directly in current CNN models without a complex model transition procedure. \cite{26} further translate the DCT representation into a sequence of DCT channel, spatial location, and DCT coefficient triples, and achieve state-of-art performance on image generation and restoration tasks with a Transformer-based auto-regressive architecture. 

The essence of the SR task is to recover the information of high-frequency channels in the image. Hence, frequency-domain features are informative for HR reconstruction and can potentially enhance the performance with proper methods. However, there is no existing SR method using the characteristics of DCT feature maps. Hence, we propose a super-resolution pipeline on the DCT domain, which we will present in detail in the 
following section. 

Precisely, the overall architecture is given in figure \ref{fig:3}. In section \ref{3.1}, we introduce the image conversion process that projects the spatial image to spatial domain. In section \ref{3.2} and \ref{3.3}, we explain the architecture and components of FreqNet in detail. The propsed frequency-domain loss function will be presented in section \ref{3.4}.

\section{Method}

\begin{figure}[H]
  \centering
  \includegraphics[width=0.95\linewidth]{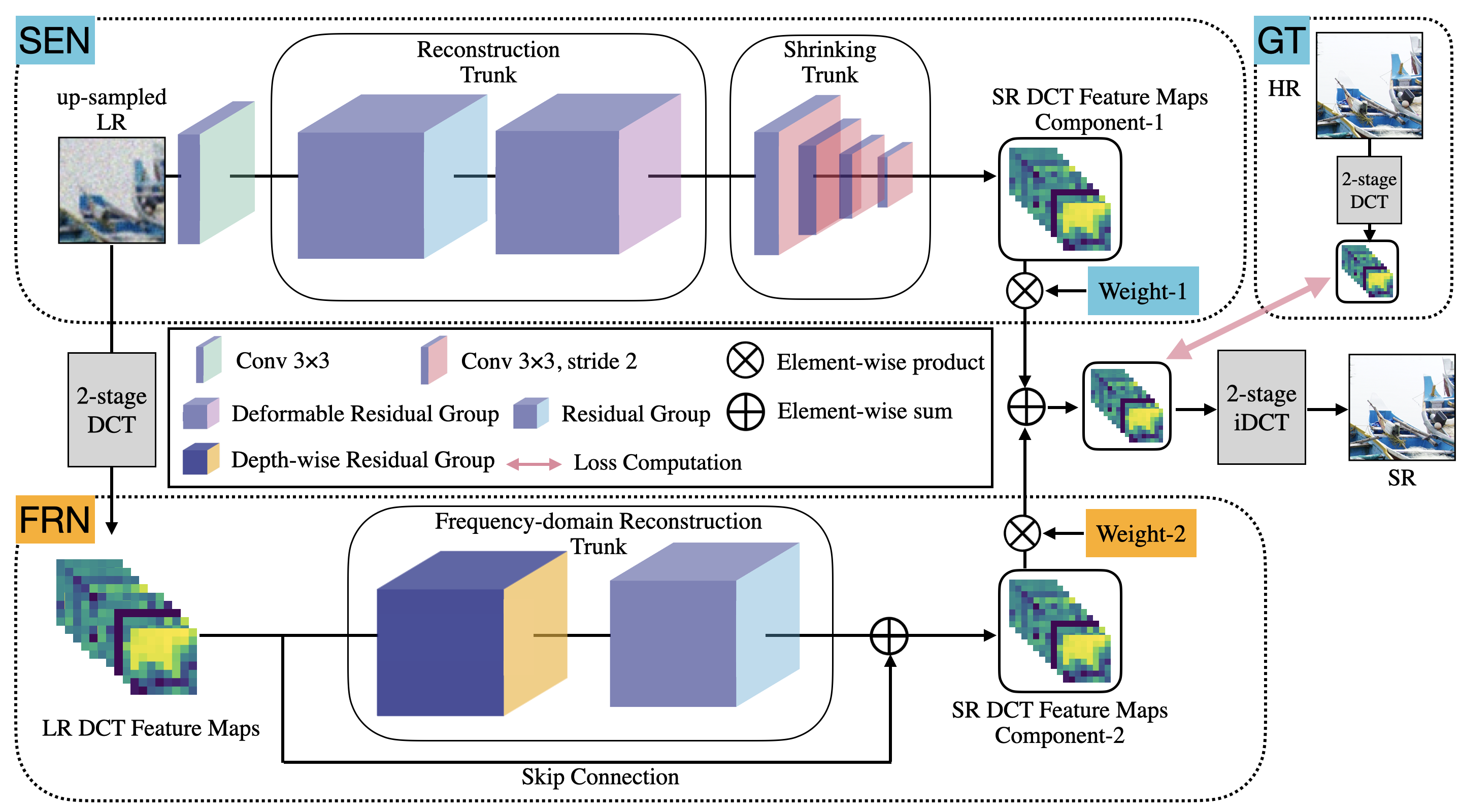}
  \hspace{1cm}
  \caption[这里将出现在插图索引中]
    {The architecture of FreqNet. Our FreqNet contains two parallel data flows: the Spatial Extraction Network(SEN) and the Frequency Reconstruction Network(FRN), taking LR image $I_{LR}$ and LR DCT feature maps as input, respectively. The final output is the weighted sum of predicted DCT feature maps from two sub-networks. Loss is computed between GT DCT feature maps(on the right-top of the figure) and the final output.}
  \label{fig:3}
\end{figure}

We propose a frequency-domain based pipeline for $4\times$ image super-resolution. Our method consists of an image conversion process that converts the spatial image to the frequency domain and a specialized network for training with the frequency domain information. As shown in \ref{fig:3}, our proposed network consists of two parallel sub-network, respectively operates on spatial-domain and frequency-domain inputs to make use of both domains’ information. We will first explain the image conversion process in the following section. Details of architecture will be discussed in \ref{3.2}.

\subsection{Image Conversion to the Frequency Domain}
\label{3.1}

Following the JPEG codec, we first transform the original RGB images to zero-centered normalized YCrCb color space, containing a brightness component Y (luma) and two color components Cb and Cr (chroma). Then we upsample the LR image to make it the same size as the HR image. 

\subsubsection{Generation of DCT Blocks}
\label{3.1.1}

\begin{figure}[h]
  \centering
  \includegraphics[width=0.95\linewidth]{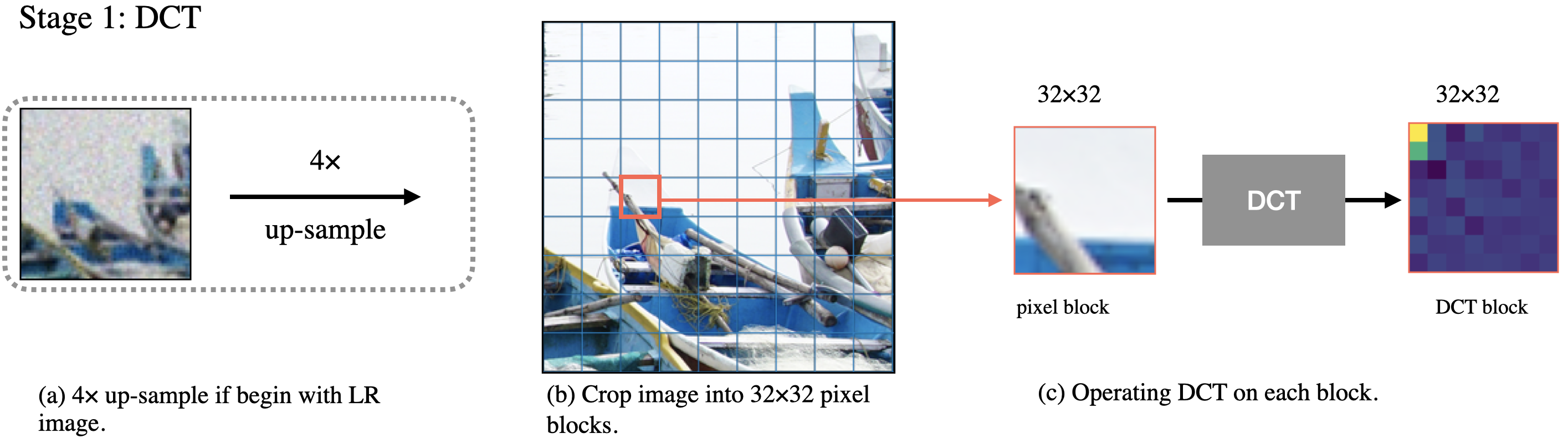}
  \hspace{1cm}
  \caption[这里将出现在插图索引中]
    {Converting pixel blocks to DCT blocks}
  \label{fig:1}
\end{figure}

To get frequency-domain information, we crop the images into uniform size of pixel blocks, then pass them through Discrete Cosine Transform(DCT) module. The DCT projects an image into a collection of cosine components which stands for different frequencies of 2D signals. Given a block size $M$, the two-dimensional DCT converts a zero-centered $M \times M$ pixel blocks $P$ to obtain an $M \times M$ DCT block $D$, as interpreted below: 
\begin{align}
    &D_{uv} = \frac{1}{4} \alpha (u)\alpha (v)\times \sum_{i=0}^{M-1}\sum_{j=0}^{M-1} P_{ij} cos(\frac{(2i+1)u\pi}{2M})cos(\frac{(2j+1)v\pi}{2M}) \\
    &\alpha(x) = \left\{
        \begin{array}{l}
        \frac{1}{\sqrt{2}} \text{ , if } x=0 \\
        1 \text{ , otherwise } 
        \end{array}
    \right. 
\label{eq1+2}
\end{align}
Where $u$ and $v$ are the horizontal and vertical index of frequencies in the DCT block, $i$ and $j$ stand for the horizontal and vertical index of pixel block, and $\alpha$ is a normalizing scale factor to enforce orthonormality. 

For a standard DCT transform in JPEG codec, the block size M is 8, which indicates that any information in an $8 \times 8$ pixel block can be represented by a linear combination of 64 2D signals. However, in $4 \times$ image super-resolution, the $8 \times 8$ block is upscaled to $32 \times 32$. Thus we perform DCT transform with the block size 32. At this stage, both the LR and HR images are converted into frequency-domain blocks that contain the DCT information of 1024 frequency channels.

\subsubsection{Reforming DCT feature maps}
\label{3.1.2}
\begin{figure}[h]
  \centering
  \includegraphics[width=0.95\linewidth]{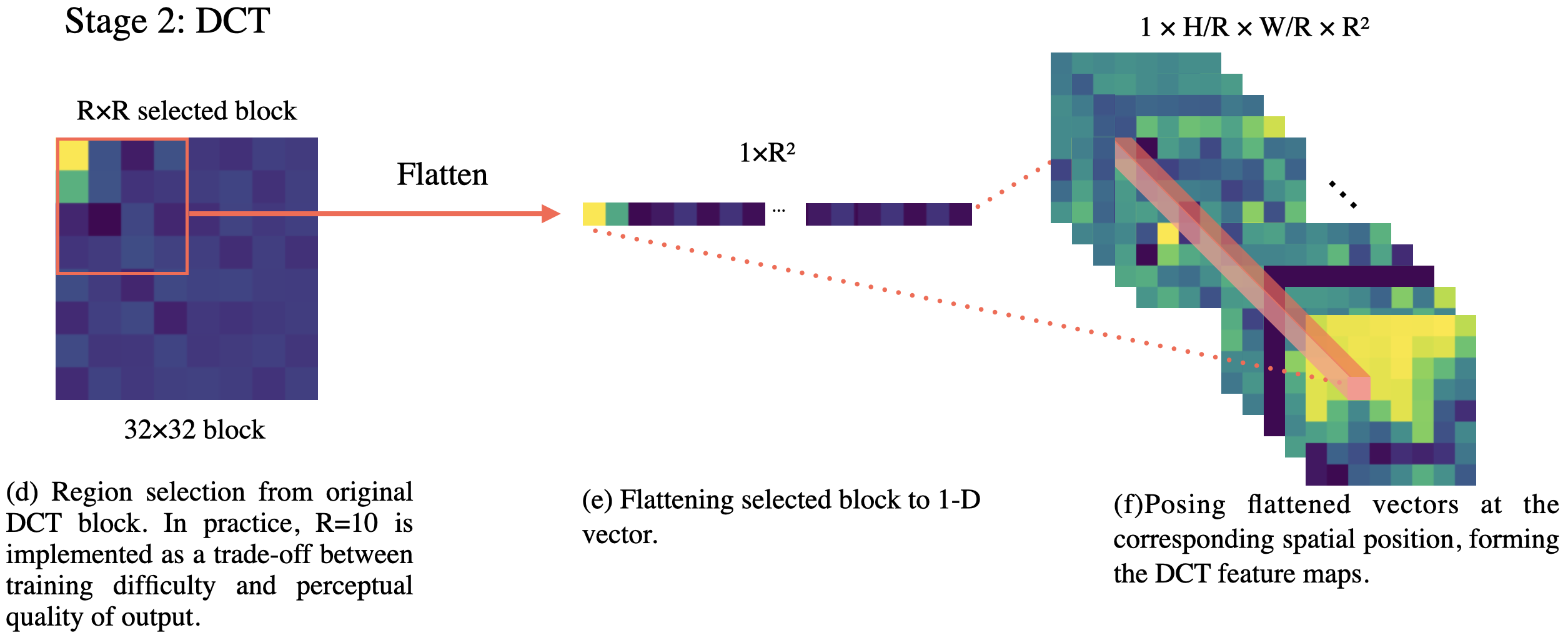}
  \hspace{1cm}
  \caption[这里将出现在插图索引中]
    {Reforming DCT blocks to DCT feature maps}
  \label{fig:2}
\end{figure}

Not all information in the $32 \times 32$ frequency range can be perceived for the perceptual ability of the human eye. Many other DCT based methods induce sparsity to DCT blocks through quantization. However, for the super-resolution task, we tend to preserve the information as much as possible. Thus, as illustrated in figure \ref{fig:2}, we perform a region selection on the DCT block. Only the values inside the left-top R×R selected region will be preserved for the next step. In practice, we choose $R=10$ as a trade-off between training difficulty and perceptual quality. We will explain later how we handle the values outside the selected region.

Following the processing method proposed by \cite{14}, we flatten the DCT blocks to DCT vectors of length $1 \times R^2$. Then we pose these vectors at their corresponding spatial positions, forming a cuboid of size $H/R \times W/R \times R^2$, where H and W are the height and width of the original image. This cuboid is a collection of DCT feature maps, each channel at the third dimension is a frequency-domain feature map that contains the information of the frequency it represents.

\subsubsection{Channel-wise Normalization}
\label{3.1.3}
We further perform normalization on each frequency channel. For channel $i$ of the frequency-domain feature maps $M$, we perform:
\begin{equation}
    M_{i_{norm}} = \frac{(M_i - Mean_i)}{Std_i}
\label{eq3}
\end{equation}
Where $Mean_i$ and $Std_i$ denotes the mean and standard deviation of channel $i$ that are pre-calculated on our training set.

Unlike quantization, this normalization process does not change the relative intensity of each feature map, thus guaranteeing the integrity of information. The purpose of this operation is to project the values to a suitable range for learning.

\subsection{Architecture of FreqNet}
\label{3.2}

As shown in figure \ref{fig:3}, our FreqNet contains two parallel data flows: the Spatial Extraction Network(SEN) and the Frequency Reconstruction Network(FRN), in order to make use of both domains' information.

The SEN takes up-scaled LR image $I_{LR}$ as input. Only one convolutional layer is used to extract the shallow feature $F_{shallow}$ from the LR input. $F_{shallow}$ is then passed through the Reconstruction Trunk(RT), which contains a sequence of multiple Residual Groups(RG)\cite{12} and Deformable Residual Groups(DRG) to convert the spatial feature maps into frequency domain features $F_{freq}$.  Then we feed $F_{freq}$ to Shrinking Trunk(ST), which consists of 4 down-sampling convolution layers with $stride=2$, to gradually shrink the scale of features maps while maintaining the channels. The final output $M_{SR_1}$ is one component of target HR DCT feature maps. The overall process can be interpreted as:
\begin{align}
 \label{eq4}
 &F_{shallow} = H_{shallow} (I_{LR}) \\
 &F_{freq} = H_{RT}(F_{shallow}) \\
 &M_{SR_1} = H_{ST}(F_{freq})
\end{align}
Where $H_{shallow}$ denotes the first convolution operation, $H_{RT}$ and $H_{ST}$ denote the RT and ST structure.

The FRN is purely operated on frequency domain. We take the pre-processed LR DCT feature maps $M_{LR}$ as input, through the frequency-domain reconstruction trunk(FRT), which contains a sequence of depth-wise residual groups(DWRG) and RG to obtain the other component of target HR DCT feature maps, noted as $M_{SR_2}$. A skip-connection is added to take advantage of the similarities between input and the target, thus drawing attention towards the difference on high-frequency channels. The overall process can be interpreted as:
\begin{align}
 \label{eq5}
 M_{SR_2} &= H_{FRT}(M_{LR}) + M_{LR}
\end{align}
Where $H_{FRT}$ denotes the FRT structure.

The outputs of two sub-network have the same size, and a weighted element-wise sum is applied to get the final output:
\begin{align}
 \label{eq6}
 M_{SR} &= M_{SR_1}\odot W_1 + M_{SR_2}\odot W_2
\end{align}
Where the $W_1$ and $W_2$ are the pre-defined weights for two components.

The output $M_{SR}$ is further fed to a 2-stage inverse Discrete Cosine Transform(iDCT) module, which is an inverse flow of data-processing pipeline we defined in \ref{3.1}. We first project the $M_{SR}$ back to its original range of values by performing denormalization on each channel. Then, in stage-1, we reform the $H/R \times W/R \times R^2$ DCT feature maps back to DCT blocks of size $R \times R$, and the rest of $32 \times 32$ block is filled with information from LR DCT blocks. Then we use iDCT to get the final SR image in stage-2.

\subsection{Modified Residual Group}
\label{3.3}
\begin{figure}[h]
  \centering
  \includegraphics[width=0.95\linewidth]{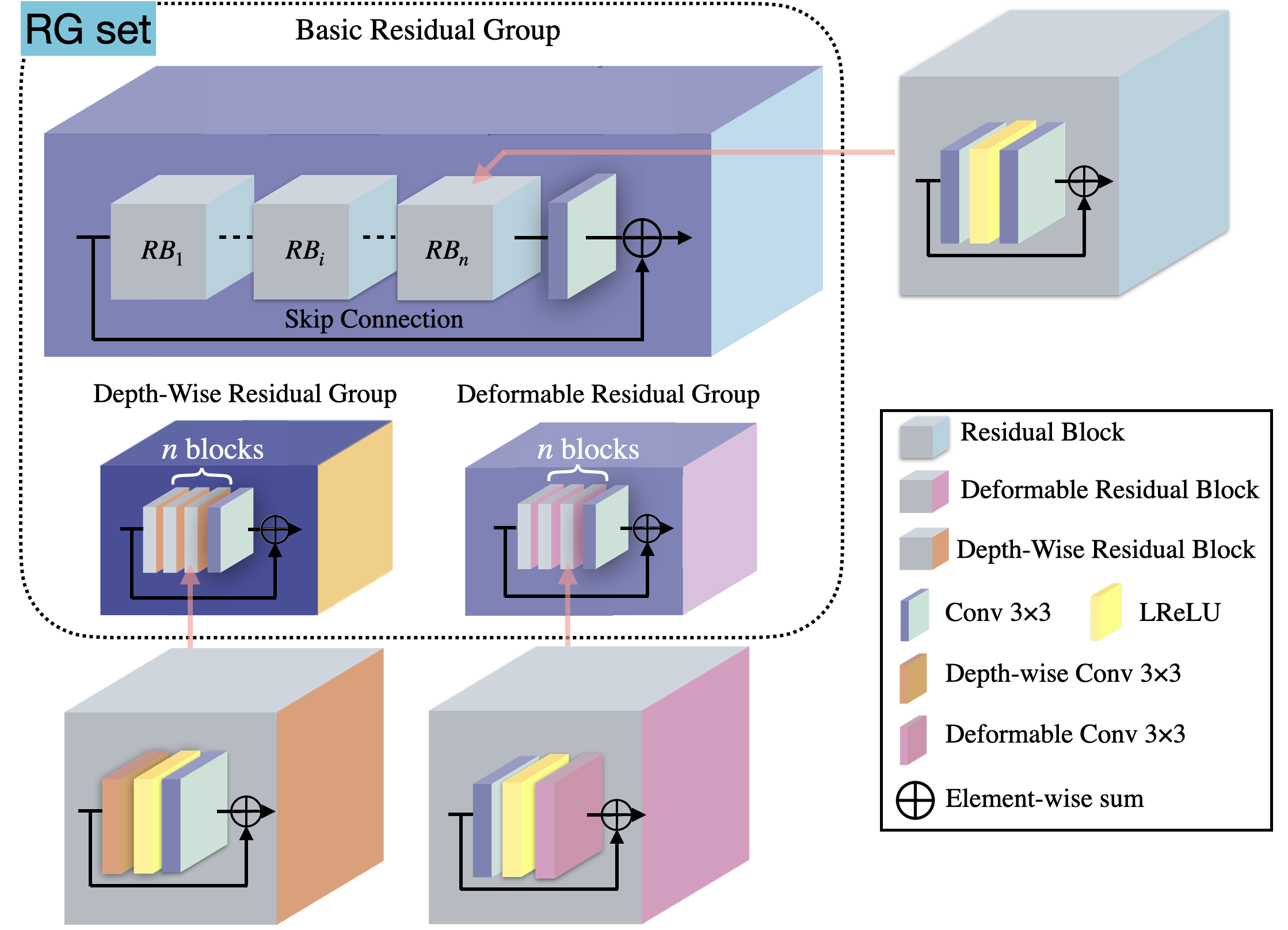}
  \hspace{1cm}
  \caption[这里将出现在插图索引中]
    {Different types of RG and RB implemented in FreqNet}
  \label{fig:4}
\end{figure}

Inspired by the success of residual groups(RG) in \cite{12}, we take it as the basic module of our network. As shown in figure \ref{fig:4}, an RG is a sequence of $n$ residual blocks(RB)\cite{8} with an in-group skip connection between the input and output features. The original RB can be interpreted as:
\begin{align}
 \label{eq7}
 &F_{res}    = H_{Conv_2}(LeakyReLu(H_{Conv_1}(F_{input}))) \\
 &F_{output} = F_{input} + F_{res}
\end{align}
Where $H_{Conv}$ denotes a convolution layer, $F_{input}$ is the feature from last block and $F_{output}$ is the feature towards next layer. As described in \ref{3.1.3}, the final output of network should be of zero-centered distribution, thus we replace ReLU layer by LeakyReLU with a high negative slope to fit our case. 

\subsubsection{Deformable Residual Group}

The RG structure makes it possible to achieve large depth, consequently providing a large receptive field size. However, uniformly extending the receptive field does not always positively impact high-precision required tasks, such as the reconstruction of frequency-domain feature maps, due to the potential redundant information. Deformable convolution layer\cite{29}(DefConv) can be a solution. By learning an offset, DefConv provides the ability to constrain the sampling area. Each convolution operation only focuses on the valuable region, reducing the impact from the redundant receptive area. 

Thus, as shown in figure \ref{fig:4} we further integrate DefConv into RB by partly replacing the original convolutional layers, introducing the deformable residual block(DRB), which is the basic module of deformable residual group(DRG):
\begin{align}
 \label{eq8}
 F_{output} &= F_{input} + H_{DefConv}(LeakyReLu(H_{Conv}(F_{input})))
\end{align}
Where $H_{DefConv}$ denotes the deformable convolution layer. The proposed DRB structure has better guidance on the receptive field, thus yield more accurate feature extraction from last layer. We implement DRG sequence in the reconstruction trunk of SEN sub-network, after a sequence of regular RG, to improve the robustness of reconstructed $F_{freq}$.

\subsubsection{Depth-wise Residual Group}

For most spatial domain tasks, the intermediate deep feature maps are abstract and strongly correlated. However, through the reforming method we defined in \ref{3.1.2}, the frequency-domain feature maps have concrete semantic information and share less correlation between each other. To better reflect this characteristic, we propose the depth-wise residual block(DWRB) that replace the first convolution layer in RB by depth-wise convolution layer\cite{28}:
\begin{align}
\label{eq9}
 F_{output} &= F_{input} + H_{Conv}(LeakyReLu(H_{DWConv}(F_{input})))
\end{align}
Where $H_{DWConv}$ denotes the depth-wise convolution layer. A depth-wise convolution layer performs 2-D convolution on each channel of the input without merging information from other channels, which is suitable to make the module focus on extracting information from own channel for the next stage of reconstruction, rather than relying on global information. Depth-wise residual group(DWRG) is the RG that deploy the DWRB instead of RB.

\subsection{Frequency-domain Loss Function}
\label{3.4}
The definition of our frequency-domain loss function $L_{freq}$ is critical to the performance of our network. Commonly, the loss function of super-resolution task is based on pixel-wise Mean Square Error(MSE), as minimizing spatial MSE also maximizes the peak signal-to-noise ratio(PSNR). However, solutions from MSE optimization can achieve high PSNR while lacking high-frequency content, which results in unsatisfying perceptual quality with overly smooth textures\cite{6}.

For our frequency domain super-resolution, this problem can be solved in a intuitive method. Since the target is a series of frequency-domain feature maps with semantic meaning assigned to each channel, we can allocate different weights to each frequency channel while computing the loss, thus explicitly guide the network to focus on the reconstruction of selected high-frequency channels. Following \cite{30}, we further replace the MSE backbone by Charbonnier Loss that can better handle the outliers, which are more likely to appear in frequency-domain samples. The proposed frequency-domain loss function $L_{freq}$ is calculated as:
\begin{align}
\label{eq10}
 &L_{Char}(x_1,x_2) = \sqrt{(x_1-x_2)^2 + \epsilon^2} \\
 &L_{freq} = \frac{1}{R^2W_{map}H_{map}} \sum_{c=1}^{R^2} \beta_c \sum_{x=1}^{W_{map}} \sum_{y=1}^{H_{map}} L_{Char}(M_{SR_{c,x,y}},M_{HR_{c,x,y}})   
\end{align}
Where $L_{Char}$ is the backbone of Charbonnier Loss, $W_{map}$ and $H_{map}$ denotes the width and height of output feature maps,  $\beta_c$ denotes the weight assigned to channel $c$ and $M$ denotes the frequency-domain feature maps, as we previously define in equation \ref{eq6}.

\section{Experiment Results}
\label{results}

\subsection{Experimental Settings}
Our experimental settings about datasets, degradation models, evaluation metric and training settings are declared below:

\textit{Datasets and degradation model.} Following \cite{27} We use 800 training images from DIV2K dataset\cite{27} as training set. For testing, we use four standard benchmark datasets: Set5\cite{33}, Set14\cite{31}, BSD100\cite{32} and MANGA109\cite{38}. We conduct experiments with Bicubic degradation model.

\textit{Evaluation Metrics.} The SR results are evaluated with PSNR on the luminance channel(Y channel) of transformed YCrCb space. We also propose a frequency-domain reconstruction metric(FRM) on the luminance channel that measures the quality of high-frequency feature reconstructed:
\begin{align}
\label{eq11}
 &FRM = 10 \ast  log_{10}(\frac{1}{L_{freq}}) 
\end{align}

\textit{Training Settings.} We crop 800 training images into mini patches. Respectively, the size of cropped LR image is $32 \times 32$ and the size of cropped HR image is $256 \times 256$. The relative location of each pair of LR and HR patches is strictly identical. Our model is trained by ADAM optimizor\cite{35}, with $\beta_1 = 0.9$, $\beta_2 = 0.99$ and $\epsilon = 10^-8$. We implement Cosine Learning Rate(CosLR) strategy, which periodically adjust the learning rate at $t_{th}$ epoch of $i_{th}$ period with the equation: 
\begin{align}
\label{eq12}
 &\eta_{i,t} = \eta_{min} + \frac{1}{2}(\eta_{max}-\eta_{min})(1 + \cos{\frac{t}{T_i}\pi}) 
\end{align}
Where the $\eta_{max}$ is $10^-4$ and $\eta_{min}$ is $10^-7$, the number of epochs in each period is 30. We use PyTorch\cite{37} to implement our method with Nvidia Geforce RTX 2080 ti GPU.

The channel-wise weights allocation of our proposed loss function $L_{freq}$(Equation \ref{eq10}) will be discussed in detail in section\ref{4.2}.

\subsection{Results with Bicubic Degradation Model}
\label{4.3}
We quantitatively compare our method with 8 State-of-the-art methods, including SRCNN\cite{3}, FSRCNN\cite{17}, EDSR\cite{8}, EDN\cite{9}, RRDB\cite{20} and its perceptual-driven method ESRGAN\cite{20}, MSRResNet\cite{6} and its perceptual-driven method MSRResNet-GAN\cite{6}. We further perform visual comparisons with these two GAN-based methods and their PSRN-oriented version to demonstrate the perceptual quality and fidelity of the output from our model.

\subsubsection{Quantitative Results by PSNR/FRM}
\label{4.3.1}
Table \ref{table3} shows quantitative comparisons for our $4\times$ SR task, we compare the average PSNR and FRM on Y channel. The PSNR results of ESRGAN and MSRResNet pair are computed using the released model. For the other models, the results are cited from their papers. All the FRM results are computed using released models. Our model has the best FRM with a slight decrease in PSNR value, which shows that our method has a more accurate reconstruction of key high-frequency information. Meanwhile, although GAN-based methods visually provide more high-frequency details, their FRM values are generally low, reflecting the lack of accuracy of high-frequency information reconstructed by such methods. We will discuss the visual behavior in detail in section \ref{4.3.2}.

\begin{table}[H]
\caption{Quantitative results with Bicubic degradation model on Y channel. Best and second best results are \textbf{highlighted} and \underline{underlined}.}
\label{table3}
\resizebox{\linewidth}{!}{
\begin{tabular}{|l|ll|ll|ll|ll|}
\hline
\multirow{2}{*}{Method} & \multicolumn{2}{l|}{Set5}            & \multicolumn{2}{l|}{Set14}           & \multicolumn{2}{l|}{Manga109}        & \multicolumn{2}{l|}{BSD100}          \\ \cline{2-9} 
                        & \multicolumn{1}{l|}{PSNR}   & FRM    & \multicolumn{1}{l|}{PSNR}   & FRM    & \multicolumn{1}{l|}{PSNR}   & FRM    & \multicolumn{1}{l|}{PSNR}   & FRM    \\ \hline
Bicubic                 & \multicolumn{1}{l|}{28.78}  & 40.06  & \multicolumn{1}{l|}{26.38}  & 39.11  & \multicolumn{1}{l|}{24.89}  & 39.65  & \multicolumn{1}{l|}{26.33}  & 38.97  \\ \hline
SRCNN                   & \multicolumn{1}{l|}{30.48}  & 40.01  & \multicolumn{1}{l|}{27.50}  & 39.09  & \multicolumn{1}{l|}{27.58}  & 39.71  & \multicolumn{1}{l|}{26.90}  & 39.11  \\ \hline
FSRCNN                  & \multicolumn{1}{l|}{30.72}  & 40.13  & \multicolumn{1}{l|}{27.61}  & 39.12  & \multicolumn{1}{l|}{27.90}  & 39.77  & \multicolumn{1}{l|}{26.98}  & 39.09  \\ \hline
MSRResNet               & \multicolumn{1}{l|}{32.22}  & 40.19  & \multicolumn{1}{l|}{28.63}  & 39.26  & \multicolumn{1}{l|}{30.48}  & 40.04  & \multicolumn{1}{l|}{27.59}  & 39.31  \\ \hline
MSRResNet-GAN           & \multicolumn{1}{l|}{29.40}  & 39.64  & \multicolumn{1}{l|}{26.02}  & 38.84  & \multicolumn{1}{l|}{27.69}  & 39.12  & \multicolumn{1}{l|}{25.16}  & 39.01  \\ \hline
EDSR                    & \multicolumn{1}{l|}{32.46}  & 40.32  & \multicolumn{1}{l|}{28.80}  & 39.61  & \multicolumn{1}{l|}{\underline{31.02}} & 40.46  & \multicolumn{1}{l|}{27.71}  & 39.25  \\ \hline
RDN                     & \multicolumn{1}{l|}{\underline{32.47}} & 40.27  & \multicolumn{1}{l|}{\underline{28.81}} & 39.47  & \multicolumn{1}{l|}{31.00}  & \underline{40.71} & \multicolumn{1}{l|}{\underline{27.71}} & 39.23  \\ \hline
RRDB                    & \multicolumn{1}{l|}{\textbf{32.60}} & \underline{40.34} & \multicolumn{1}{l|}{\textbf{28.88}} & \underline{40.14} & \multicolumn{1}{l|}{\textbf{31.16}} & 40.63  & \multicolumn{1}{l|}{\textbf{27.76}} & \underline{39.52} \\ \hline
ESRGAN                  & \multicolumn{1}{l|}{29.56}  & 39.38  & \multicolumn{1}{l|}{26.19}  & 38.79  & \multicolumn{1}{l|}{28.03}  & 39.28  & \multicolumn{1}{l|}{25.32}  & 38.86  \\ \hline
FreqNet(Ours)           & \multicolumn{1}{l|}{32.08}  & \textbf{43.56} & \multicolumn{1}{l|}{28.47}  & \textbf{42.60} & \multicolumn{1}{l|}{30.23}  & \textbf{40.91} & \multicolumn{1}{l|}{27.51}  & \textbf{40.87} \\ \hline
\end{tabular}}
\end{table}

\subsubsection{Visual Results}

\begin{figure}[H]
    \centering
    
    \begin{subfigure}[b]{.47\linewidth}
    \includegraphics[width=\linewidth]{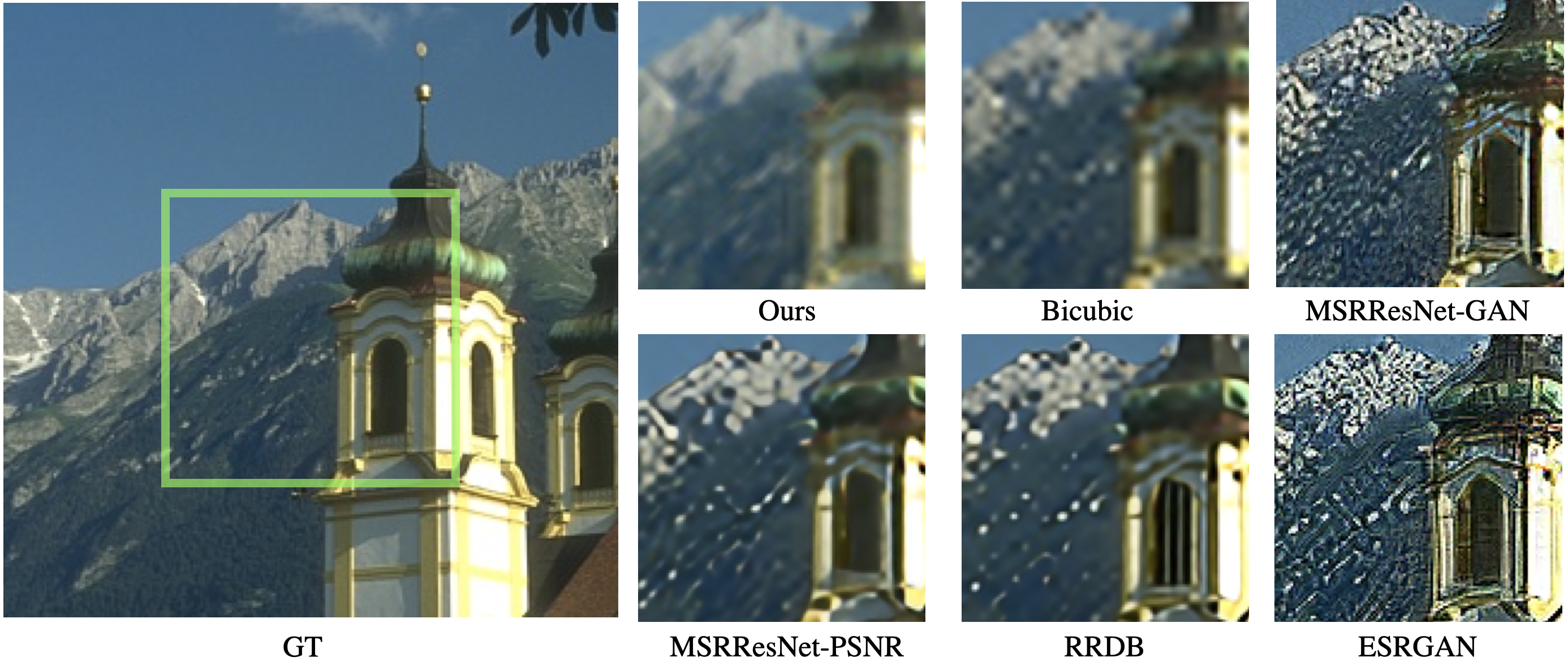}
    \caption{Image ``126007.png"}\label{fig:res126007}
    \end{subfigure}\hfill
    \begin{subfigure}[b]{.47\linewidth}
    \includegraphics[width=\linewidth]{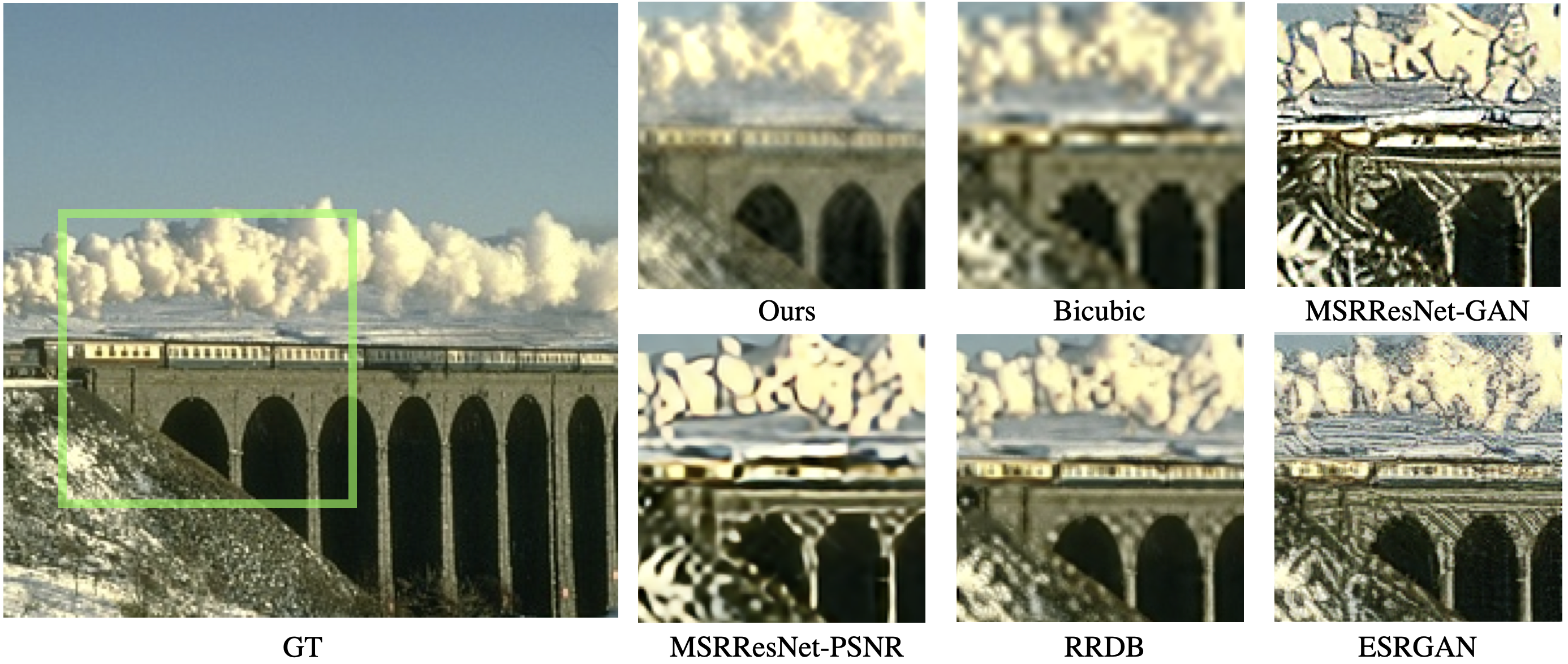}
    \caption{Image ``182053.png"}\label{fig:res182053}
    \end{subfigure}
    
    \begin{subfigure}[b]{.47\linewidth}
    \includegraphics[width=\linewidth]{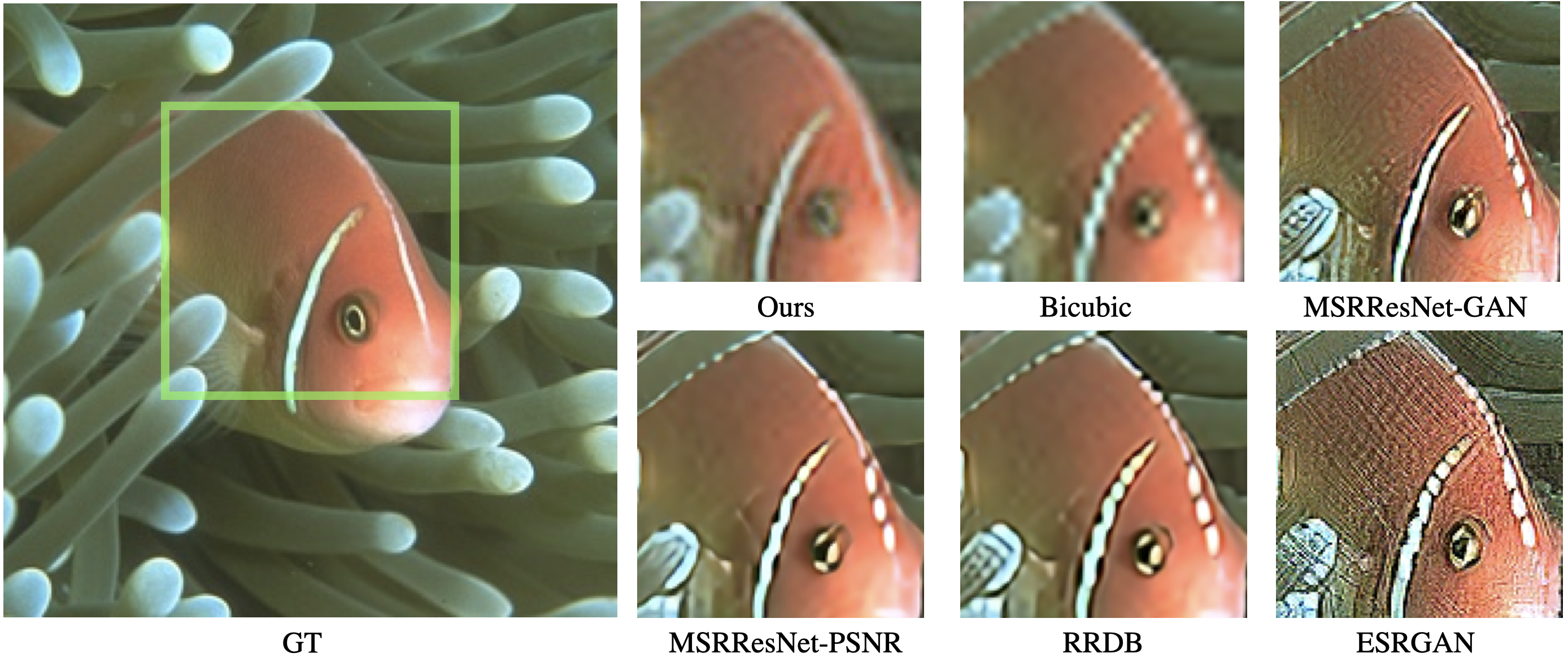}
    \caption{Image ``210088.png"}\label{fig:res210088}
    \end{subfigure}\hfill
    \begin{subfigure}[b]{.47\linewidth}
    \includegraphics[width=\linewidth]{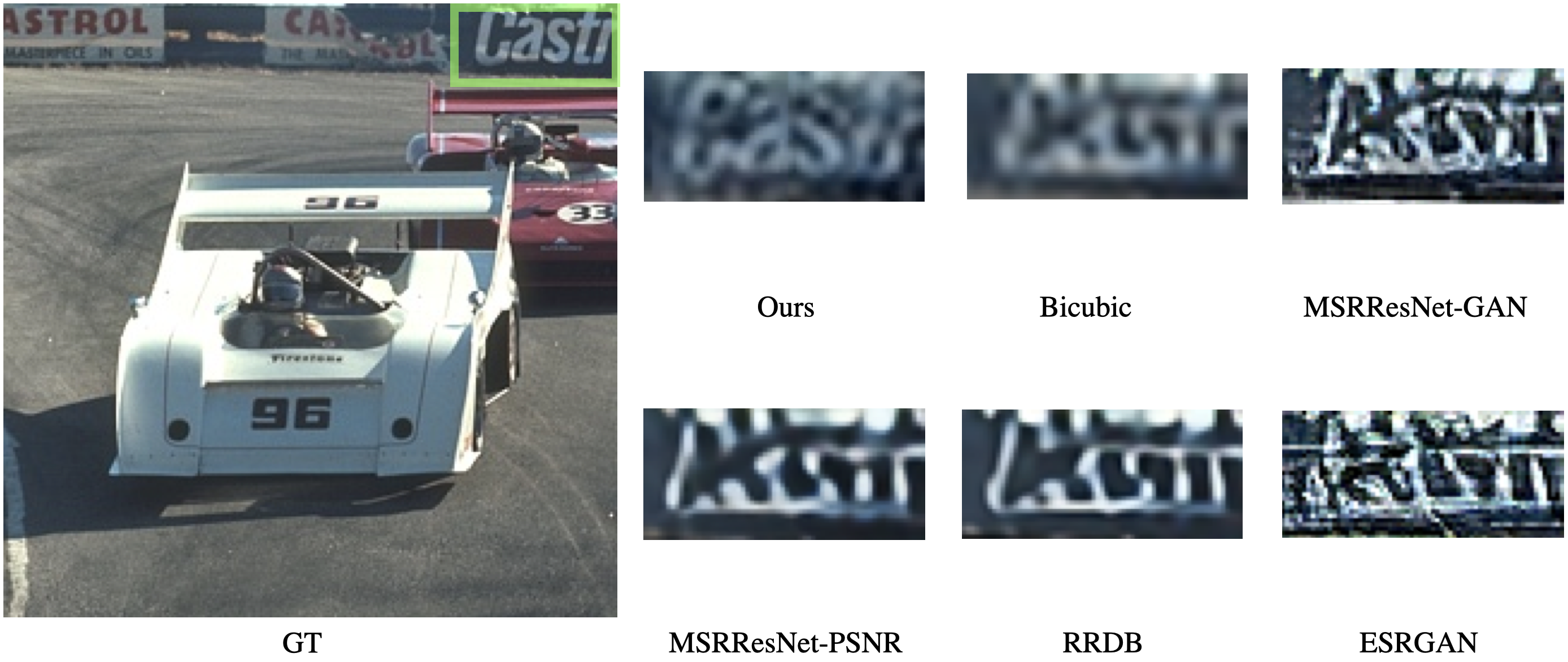}
    \caption{Image ``21077.png"}\label{fig:res21077}
    \end{subfigure}
    
    \begin{subfigure}[b]{.95\linewidth}
    \includegraphics[width=\linewidth]{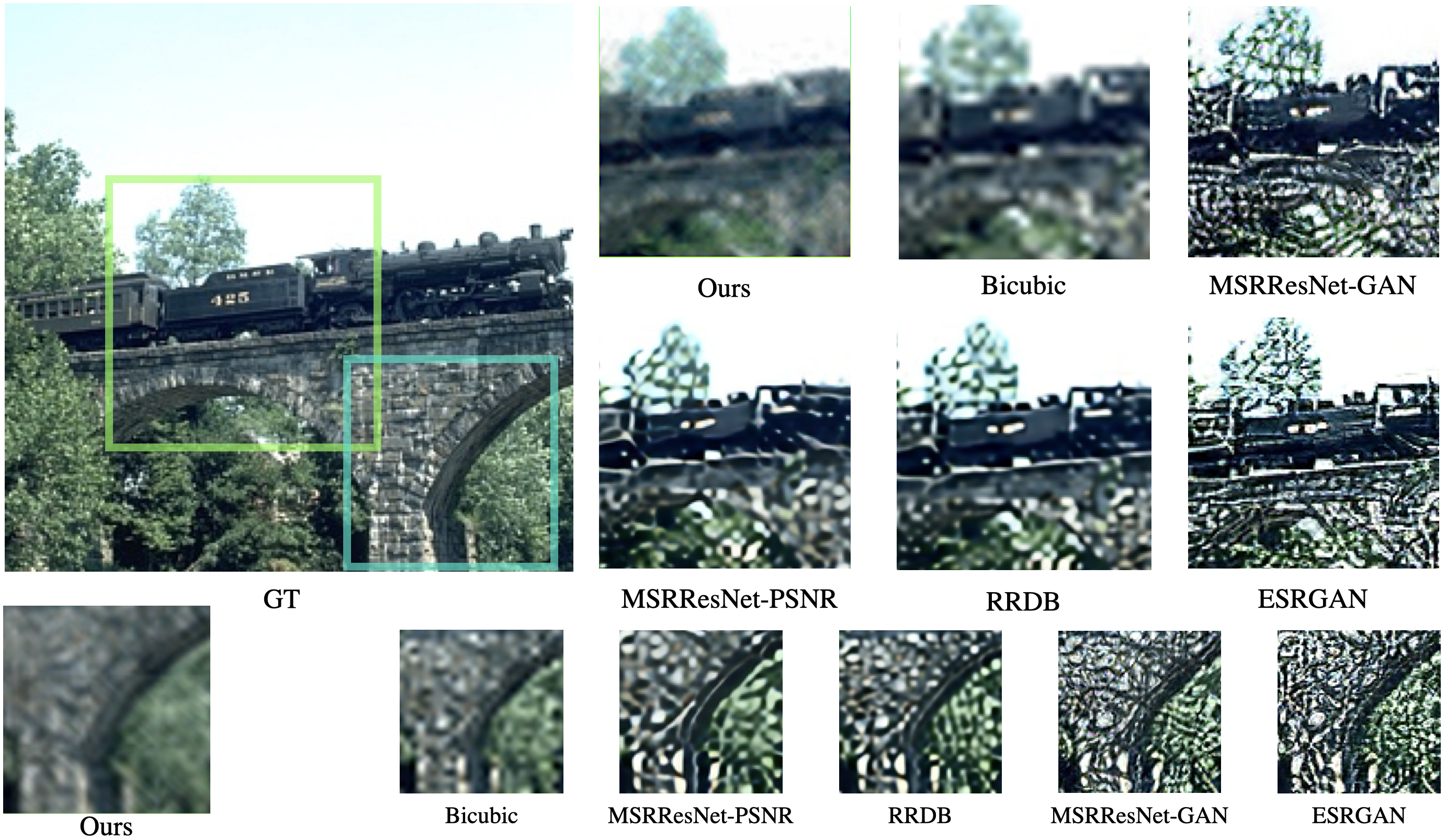}
    \caption{Image ``351093.png"}\label{fig:res351093}
    \end{subfigure}
    
    \caption[这里将出现在插图索引中]
    {Visual comparison for 4$\times$ SR with Bicubic Degradation model on BSD100 datasets.}
    \label{fig:visualize}
\end{figure}

\label{4.3.2}
In figure \ref{fig:visualize}, we show visual comparisons of SR results with the Bicubic Degradation model on BSD100 datasets. For images ``126007.png" and ``351093.png", we observe that our method has more precise building contours than the PSNR method, contains more details, and does not have the excessive texture as in the GAN method. For image ``210088.png", we observe that our method produces the best face pattern and eye details for the clownfish. For image ``21077.png", our method better restores the text ``cas" over the other methods. And in image ``182053.png", our method predicts the arches correctly while having fewer unnecessary artifacts. 

\subsection{Effects of Frequency-domain Loss Function and Modified RG}
\label{4.2}
We study the effects of proposed Deformable Residual Group, Depth-wise Residual Group and the Frequency-domain Loss Function.

\subsubsection{Settings and Effects of Frequency-domain Loss Function.} 

\begin{figure}[h]
  \centering
  \includegraphics[width=0.95\linewidth]{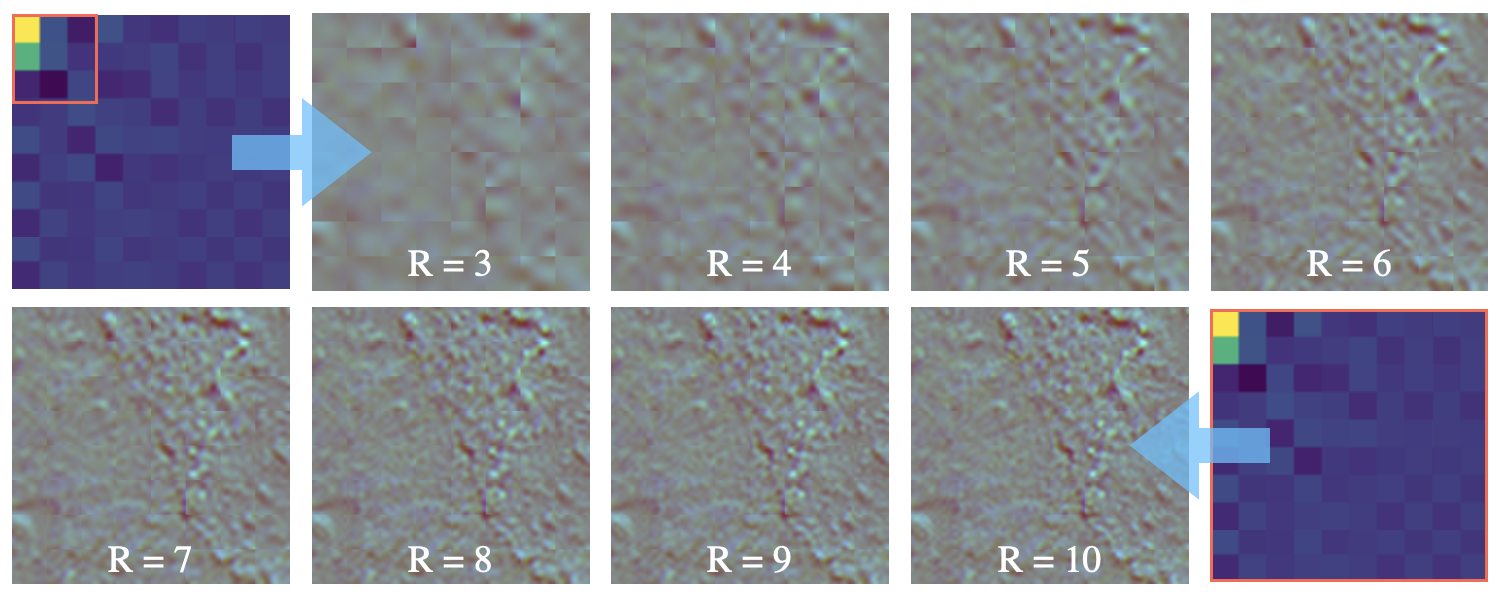}
  \hspace{1cm}
  \caption[这里将出现在插图索引中]
    {Progressively calculate residuals between HR and LR pixel-blocks under different size of region selection.}
  \label{fig:5}
\end{figure}

As we defined in Equation \ref{eq10}, each channel has a pre-assigned weight. We propose a statistical solution to decide the weight of each channel coarsely. As shown in Figure \ref{fig:5}, given a pair of HR and up-sampled LR DCT blocks, after the region selection process(i.e. Figure \ref{fig:2}, (d)), of size $R \times R$, we perform 8 times of computation in total. For the $i\text{-th}$ computation, we keep the $(i+2) \times (i+2)$ DCT region at the left-top of original DCT block unchanged, and set the values outside the selected region to be $0$. Then we perform iDCT on both LR and HR DCT blocks and compute the mean pixel-wise residual $res_i$ of two converted pixel blocks as:
\begin{align}
\label{eq13}
 &res_i = \left| \frac{I_{HR} - I_{LR}}{R^2} \right|
\end{align}

\begin{table}[h]
\centering
\caption{Weight Allocation.}
\label{table1}
\begin{tabular}{l|l|l|l|l|l|l|l|l|}
\cline{2-9}
Region: &3 & 4-3 & 5-4 & 6-5 & 7-6 & 8-7 & 9-8 & 10-9 \\ \cline{2-9} 
\cline{2-9} 
$\beta_c$           & 1   & 1     & 5     & 10    & 10    & 5     & 1     & 1      \\ \cline{2-9} 
\end{tabular}
\end{table}

We randomly picked 1000 samples from the training set to perform the statistics by accumulating $res_i$. We define $res_0 = 0$, then for each $i$, the value $v_i = res_i - res_{i-1}$ reflects the difference between HR and LR images while considering the addition frequency channels of $R=i+2$, which is proportional to their importance. Therefore, based on $[v_i]_{i\in[1,8]}$, we allocate weights as the table \ref{table1} shows, where Region $i-(i-1)$ denotes the additional channels between the left-top $i \times i$ region and $(i-1) \times (i-1)$ region of the $R \times R$ DCT block, and $\beta$ denotes the weight assigned to these channels involved in Equation \ref{eq10}.

\begin{figure}[H]
  \centering
  \includegraphics[width=0.95\linewidth]{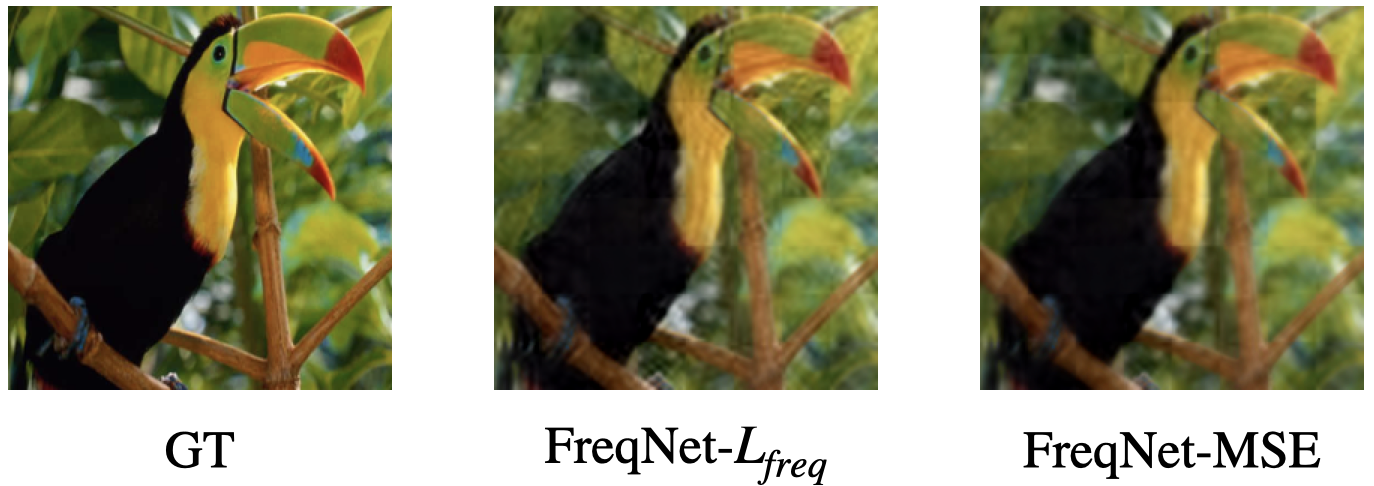}
  \hspace{1cm}
  \caption[这里将出现在插图索引中]
    {Comparison of MSE-supervised and $L_{freq}$-supervised FreqNet. $L_{freq}$ supervised result contains more high-frequency details.}
  \label{fig:7}
\end{figure}

To demonstrate the effect of the proposed frequency-domain loss function $L_{freq}$, we run the training process with MSE and $L_{freq}$ respectively, and compare the output of two models on Set5. Both the PSNR and FRM of $L_{freq}$ supervised model is higher than the MSE supervised model, and the output images contain more accurate high-frequency texture. Figure \ref{fig:7} shows the comparison of the SR results of image ``bird.png" between MSE-supervised and $L_{freq}$-supervised FreqNet after same number of iterations. The $L_{freq}$-supervised model can produce more high-frequency details.

\subsubsection{Effects of Deformable(DRG) and Depth-wise Residual Group(DWRG).} 

We perform a series of ablation experiments by replacing DRG or/and DWRG with original RG, to demonstrate the effect of our modified RG structure. 

\begin{table}[h]
\centering
\caption{Ablation Experiments on DWRG and DWG. We use PSNR and our proposed FRM as the metric.}
\label{table2}
\begin{tabular}{|l|l|l|l|l|}
\hline
DRG         &\ding{55}  &\checkmark &\ding{55}  &\checkmark  \\ \cline{1-1}
DWRG          &\ding{55}  &\ding{55}  &\checkmark &\checkmark  \\ \hline
PSNR on Set5 & 31.88 & 32.06 & 31.91 & \textbf{32.08} \\ \hline
FRM on Set5  & 43.24 & 43.51 & 43.29 & \textbf{43.56} \\ \hline
\end{tabular}
\end{table}

Respectively, in Spatial Extraction Network(SEN) we set $num_{DRG} = 3$ and $num_{RG} = 7$, in Frequency Reconstruction Network(FRN) we set $num_{DWRG} = 3$ and $num_{RG} = 7$. For each group, the number of residual blocks is set as 10. As shown in Table \ref{table2}, the PSNR on Set5 increased by 0.18 dB when we replace specific RG with DRG, increased by 0.03 dB when we replace specific DWRG, and we can have the best performance by using both of them. The FRM on Set5 also increased when we replace RG with DRG and DWRG, by 0.27 and 0.05 respectively. The comparison shows the effectiveness of our proposed modified Residual Group architectures.

\subsection{High-frequency Detail Enhancement based on other SR Models}

As the output of our proposed model is a group of separated frequency-domain feature maps, we can easily merge our output with the output of other SR models, thus realize the enhancement on selected high-frequency channels. We first perform a similar process as \ref{3.1} to convert the output from other SR model $I_{SR_ori}$ to its frequency-domain feature maps group $F_{SR_ori}$, then we replace certain channels in $F_{SR_ori}$ with the corresponding channels in $F_{output}$ from FreqNet to get the merged output $F_{merge}$.

\begin{figure}[H]
  \centering
  \includegraphics[width=0.95\linewidth]{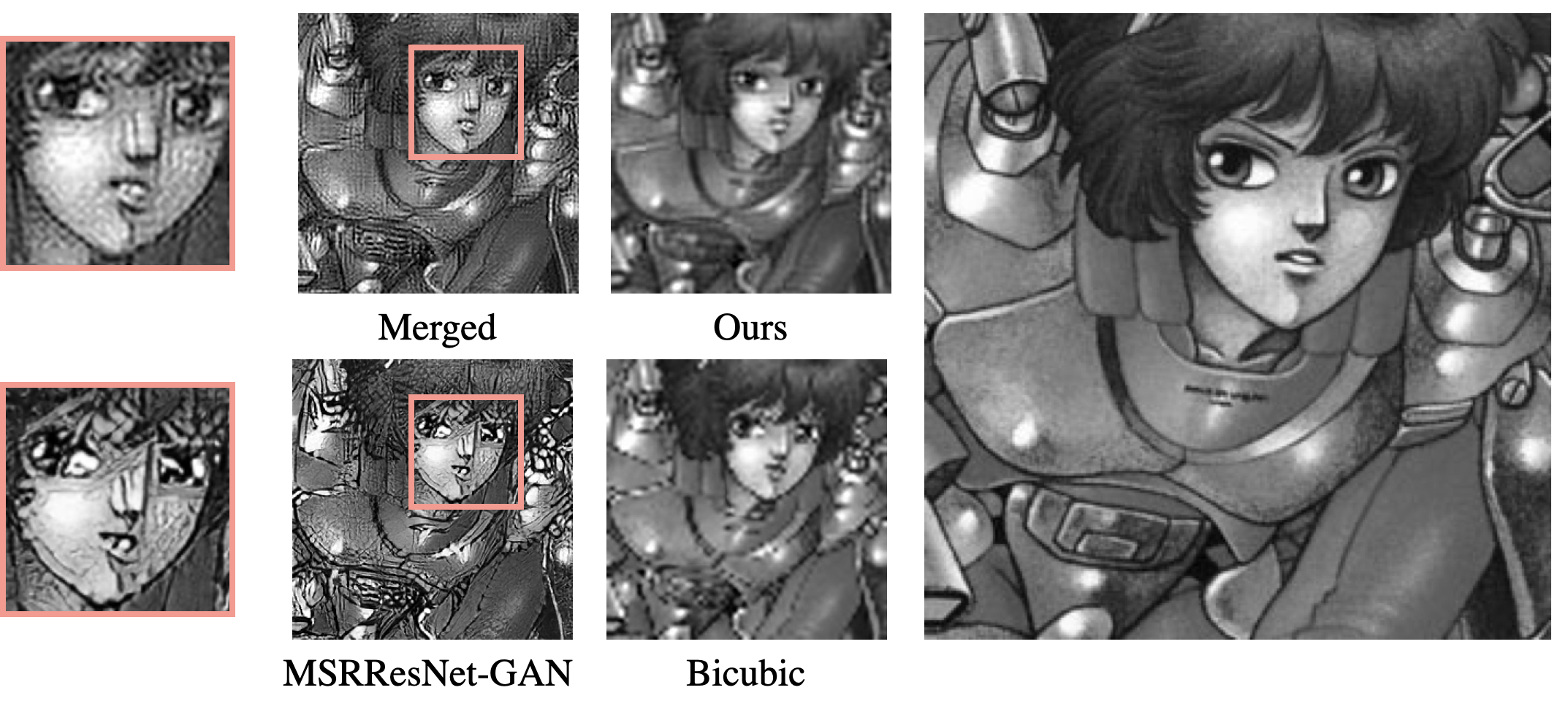}
  \hspace{1cm}
  \caption[这里将出现在插图索引中]
    {Merging with the output of FreqNet can reduce unreasonable artifacts from the GAN method while maintaining the details in the picture.}
  \label{fig:6}
\end{figure}

Specifically, we can merge our output with GAN-based SR models. As shown in Figure \ref{fig:6}, we merge the output of MSRResNet-GAN\cite{20} with the output of our model, for image ``ARMS.png" in ``Manga109"\cite{38}, the results are presented in Y-channel. The excessive artifacts from GAN can be corrected by channel replacement, and the reasonable high-frequency information that doesn't ruin the fidelity can be preserved. This method is practical when the output is blurred due to the difficulty of prediction.

\section{Conclusions}

We propose FreqNet, a frequency-domain image super-resolution model that explicitly learn the reconstruction of high-frequency details from LR images. We propose the depth-wise residual group(DWRG) and deformable residual group(DRG) structure to fit the characteristics of frequency-domain task and improve the ability of our network. Meanwhile, we propose a frequency-domain loss function and the frequency-domain reconstruction metric(FRM) that can measure the quality of high-frequency detail reconstruction. The quantitative and visual results demonstrate the effectiveness of our method, and we can further merge the output of our network with the other SR models as a post-processing enhancement. 

% \section*{Declaration}

% The authors declare that they have no known competing financial interests or personal relationships that could have appeared to influence the work reported in this paper.

% \section*{Acknowledgement}

% This paper is supported by 

\appendix

 \bibliographystyle{elsarticle-num} 
 \bibliography{cas-refs}





% \end{thebibliography}
\end{document}